\documentclass[aps,prd,12pt,eqsecnum, %showpacs,
preprintnumbers,nofootinbib,superscriptaddress]{revtex4}
\usepackage{amssymb,amsmath,amsthm,graphicx,amscd,color,mathtools}
\usepackage{upgreek,enumerate,color,verbatim,multirow,comment,ulem,slashed}
\usepackage[hidelinks]{hyperref}

\newcommand{\llangle}{\langle \hspace{-.8mm}\langle}  
\newcommand{\rrangle}{\rangle \hspace{-.8mm}\rangle}
\makeatletter                    %allow @ in command names
\@addtoreset{equation}{section}  %reset equation count in each section
\makeatother                     %disallow @ in command names

\begin{document}

\title{Fluctuation-Dissipation and Correlation-Propagation Relations in $(1+3)$D  Moving Detector-Quantum Field Systems}

%\author{J.-T. Hsiang\corref{cor1}} \ead{cosmology@gmail.com}\address{Center for High Energy and High Field Physics, National Central University, Chungli 32001, Taiwan} \author{B. L. Hu} \ead{blhu@umd.edu} \address{Maryland Center for Fundamental Physics and Joint Quantum Institute, University of Maryland, College Park, Maryland 20742-4111, USA} \author{S.-Y. Lin} \ead{2@aol.com} \address{Department of Physics, National Changhua University of Education, Changhua 50007, Taiwan} \author{K. Yamamoto} \ead{3@aol.com} \address{???, Kyushu University, Japan} \cortext[cor1]{Corresponding author}

\author{Jen-Tsung Hsiang}
\email{cosmology@gmail.com}
\affiliation{Center for High Energy and High Field Physics, National Central University, Chungli 32001, Taiwan}
\author{B. L. Hu}
\email{blhu@umd.edu}
\affiliation{Maryland Center for Fundamental Physics and Joint Quantum Institute, University of Maryland, College Park, Maryland 20742-4111, USA}
\author{Shih-Yuin Lin}
\email{sylin@cc.ncue.edu.tw}
\affiliation{Department of Physics, National Changhua University of Education, Changhua 50007, Taiwan}
\author{Kazuhiro Yamamoto}  
\email{yamamoto@phys.kyushu-u.ac.jp}
\affiliation{Department of Physics, Kyushu University, Fukuoka 819-0395, Japan}

\date{\small \today}

\begin{abstract}
The fluctuation-dissipation relations  (FDR) are powerful relations which can capture the essence of  the interplay between a system and its environment. Challenging problems of this nature which FDRs aid  in our understanding include the backreaction of quantum field processes like particle creation on the spacetime dynamics in early universe cosmology or quantum black holes.  
%The detector-field model is a simplified yet useful system for investigating these relations, such as carried out by Raval et al \cite{RHA}, who discovered the 
The less familiar, yet equally important correlation-propagation relations (CPR) relate the correlations of stochastic forces on different detectors to the retarded and advanced parts of the radiation propagated in  the field.  Here, we analyze a system of $N$ uniformly-accelerated Unruh-DeWitt detectors whose internal degrees of freedom (idf) %$Q_i$  
are minimally coupled to a real, massless, scalar field %$\phi$ 
in 4D Minkowski space, extending prior work in 2D with derivative coupling. %\cite{CPR2D}.
Using the influence functional formalism, we derive the stochastic equations describing the nonequilibrium dynamics of the idfs.    We show after the detector-field dynamics has reached equilibration the existence of the FDR and the CPR %relations 
for the detectors, which combine to form a {\it generalized} fluctuation-dissipation matrix relation  We show explicitly the energy flows between the constituents of the system of detectors and between the system and the quantum field environment. This power balance anchors the generalized FDR.  We anticipate this matrix relation to provide a useful guardrail in expounding some basic issues in relativistic quantum information, such as ensuring the self-consistency of the energy balance and tracking the quantum information transfer in the detector-field system.
\end{abstract}

%\begin{keyword} fluctuation-dissipation relation \sep correlation-propagation relations \sep Unruh-DeWitt detector \sep uniform acceleration \end{keyword}

\maketitle

\baselineskip=18pt
\allowdisplaybreaks

%\newpage
\section{Introduction}

Fluctuation-dissipation relations  (FDR), though rooted in statistical mechanics \cite{Kubo,FetWal}, has wide-ranging implications and applications. For example,  it captures the essence of the so-called backreaction problem in particle-field systems and in 
%gravitation-cosmology.  
{gravitational and cosmological physics.} 
Sciama \cite{Sciama} treated black holes with Hawking radiation \cite{Haw75} as a dissipative system,  and, with Candelas, proposed  to view its interaction with a quantum field  in the light of a FDR \cite{CanSci77} 
{(see also Mottola \cite{Mot86}).  Hu \& Sinha, Campos \& Verdaguer} %{\it et al} 
\cite{HuSin95,CamVer96,CamHu98} showed how the backreaction of particle creation on the geometrodynamics of the early universe can be phrased in terms of a FDR.  

The existence of FDR in a thermal field in the context of linear response theory (LRT) \cite{Kubo,FetWal} is better known than the correlation-propagation relations (CPR), the existence of which in a quantum field was first discovered by Raval, Hu and Anglin. In Ref.~\cite{RHA} they showed that there exists 1) a set of {FDR} %{\it fluctuation-dissipation relations} 
relating the fluctuations of the stochastic forces to the dissipative forces on the detectors, and 2) a related set of {CPR} %\textit{correlation-propagation relations} 
between the correlations of stochastic forces on different detectors and the retarded and advanced parts of the radiation mediated by them.

{Here the detector is a physical object with internal degrees of freedom, like an atom.
A uniformly accelerated detector moving in a quantum field  was used by Unruh \cite{Unr76} to illuminate the physics of Hawking effect. Considering a detector's motion and interaction with the quantum field, the detector-field system  opens up  accessible channels to probe  experimentally into the  many properties of a moving detector or atom in a quantum field, such as quantum radiatio\cite{SCD,Tak,GinFro,Grove,RSG,Unr92,MPB,RHA,HuRavCapri,JH05,LinHu06,HIUY} and nonequilibrium, nonMarkovian effects \cite{RHK97,HJCapri,LCH08,DLBM,Lin17}.  It has also become  popular in the newly emergent field of relativistic quantum information  \cite{HLL} for tackling  environmental influences on quantum coherence  and entanglement (e.g., \cite{LinHu08,LCH08,LinHu09,LinHu00,OLMH} and references therein).}

In a recent paper \cite{CPR2D} three of the present authors considered $N$ uniformly-accelerating Unruh-DeWitt detectors \cite{Unr76,DeW79} whose internal degrees of freedom are coupled to a massless scalar field in (1+1)D Minkowski space and derived a set of  %fluctuation-dissipation relations  
{FDR and  %correlation-propagation relations 
CPR} between the dissipation and the noise kernels of the {\it detectors} after the detector-field system has reached equilibrium. Together these two sets of relations form a matrix relation, which we called the {\it generalized FDR}, with the FDR entering as the diagonal components and the CPR entering the off-diagonal components of the generalized FDR. %matrix relation.  
Though not known before, but as we shall show, the CPR is indispensable in keeping the energy balance in the detector-field system. This aspect is completely absent in the conventional derivation and understanding of FDR based on LRT because the bath used there is only a {passive-parametric rather than % and not  
an} active-dynamical quantity. The CPR contains the information about the state of motion of the detectors even when they are remotely located and have no direct coupling with one another. The ambient quantum field relays this information from one detector to another and the CPR shows how this information is propagated and how their correlations evolve. For this reason we anticipate the CPR will play an important bookkeeping role in quantum information flow and the energy balance in the detector-field system.    

In this paper we derive the FDR and the CPR %relations 
for {a similar detector-field} %the same 
system but in (1+3) dimensions, namely,  $N$ uniformly-accelerated Unruh-DeWitt detectors whose internal degrees of freedom $Q_i$  are coupled to a real, massless, scalar field $\phi$ in 4D Minkowski space. The major difference {from the 2D case} 
%between these two cases 
is in the type of coupling between the detectors and the field. We first write down the Langevin equation governing the internal degrees of freedom of the $i$-th detector interacting with the quantum field and with that of the other $N-1$ detectors mediated by the field.  We  explain the properties of the two Green functions appearing therein: the retarded Green function representing dissipation and the Hadamard function, the noise, as a coarse-grained representation of the fluctuations in the quantum field.  We show the existence of a generalized FDR  after the detector-field dynamics (of the internal degrees of freedom of the $N$ detectors interacting with a quantum field) has reached  equilibration. We then proceed to  establish the FDR and CPR for the detectors.  We comment on the important differences between the FDR derived here via nonequilibrium dynamics \cite{KadBay62,KuboBook,CalHu88,CalHu08} and that from LRT.  Next we show explicitly the energy flows between the constituents of the system of detectors and between the system and the quantum field environment. This power balance,  preconditioned on the existence of an equilibrium state, is the physical embodiment of the generalized FDR. We conclude with a note on the importance of including non-Markovian effects in the dynamics to guarantee full self-consistency.

\section{Uniformly accelerating Unruh-DeWitt detectors in (1+3)D Minkowski space}

Suppose the  detectors are at different transverse spatial locations $\mathbf{x}_{\perp}$ in the same Rindler wedge $R$  ($\lvert t\rvert<z$). They all  accelerate uniformly  in the $z$ direction from past null infinity to future null infinity  in 1+3 dimensional Minkowksi space. We further assume that at $t=0$, when the detector-field coupling of strength $\lambda$ is turned on, all detectors comes to the point  closest to the origin\footnote{This requirement is only for the sake of mathematical simplicity in the subsequent discussions. It can be relaxed to the case that the detectors may arrive at the mid-points at different times since we are interested only in late-time dynamics. The complexity of the transient regime does not concern us.}. In other words, the external degree of freedom of each detector will follow a worldline described by a certain fixed set of $\xi$ and $\mathbf{x}_{\perp}$ coordinates  in Rindler space with  line element of  the form
{
\begin{equation}\label{E:krbds}
	ds^{2}=e^{2\mathsf{a}\xi}\bigl(-d\eta^{2}+d\xi^{2}\bigr)+d\mathbf{x}_{\perp}^{2}. %+dz^{2}\,.
\end{equation}
The Rindler time $\eta$ and spatial coordinate $\xi$ are related to the Minkowski time $t$ and spatial coordinate $z$ by
\begin{align}
	t&=\frac{e^{\mathsf{a}\xi}}{\mathsf{a}}\,\sinh\mathsf{a}\eta\,,&z&=\frac{e^{\mathsf{a}\xi}}{\mathsf{a}}\,\cosh\mathsf{a}\eta\,.
\end{align}
We call the constant parameter $\mathsf{a}/2\pi$ the fiducial temperature \cite{Unr76, CPR2D}.}

The internal degree of freedom $Q_i$ of each detector is modeled by a harmonic oscillator. Its interaction term with the scalar field  takes the form~
\begin{align}
	S_{\textsc{int}}[Q,\phi]&=\int\!d^{4}x\sqrt{-g}\;j(x)\phi(x)\,,&j(x)&=\sum_{i}\lambda\int\!d\tau_{i}\;Q_{i}(\tau_{i})\,\frac{\delta^{(4)}[x^{\mu}-z_{i}^{\mu}(\tau_{i})]}{\sqrt{-g}}\,,
\end{align}
instead of the derivative coupling\footnote{If we choose a coupling of the derivative type in 4D,  the self force in the equation of motion will depend on the third-order time derivative of $Q$, which unravels pathological behaviors associated with singular differential equation~\cite{Jackson, Rohlich,Yaghjian}. A common remedy is to order-reduce~\cite{Roh00} the damping term to first-order time derivative, based on the concept of the critical manifold~\cite{Spohn}. However, the order-reduced dynamics will result in an FDR  where the proportionality factor depends on the parameters of the reduced systems, as well as on those of the environment.  This is not what a generic, categorical relation such as FDR should look like.} in the (1+1)D case. Here $z_{i}^{\mu}$ traces the prescribed worldline of the $i^{\text{th}}$ detector,  $\tau_{i}$ is its proper time and $g$ is the determinant of the metric \eqref{E:krbds}. The detector-field coupling is the same as in the model  studied in great details in~\cite{QTD1}, except that the detectors here are uniformly accelerating.  For economy of space we refer many technical details to that paper.

\subsection{Langevin equation, Retarded Green function and Hadamard function}

Using the Feynman-Vernon ~\cite{FeyVer63} influence functional~\cite{CalLeg83,GSI88,HPZ92,HHAoP} and the related Schwinger-Keldysh \cite{Sch61,Kel63} closed-time-path ~\cite{Chou85} or `in-in' \cite{DeW86,Jor86,CalHu88} formalism, we coarse-grain the field degrees of freedom in the total action of the detector-field system to derive a stochastic effective action. Variation of it gives a stochastic equation of motion, the Langevin equation,  for the internal degrees of freedom of each detector 
\begin{equation}\label{E:bfsrdf1}
	m\ddot{Q}_{i}(\tau_{i})+m\omega^{2}\,Q_{i}(\tau_{i})-\lambda^{2}\sum_{j}\int\!d\tau_{j}\;\Bigl\{G_{R}^{(\phi)}[\tau_{i},\mathbf{z}_{i}(\tau_{i});\tau_{j},\mathbf{z}_{j}(\tau_{j})]\Bigr\}_{ij}\,Q_{j}(\tau_{j})=\lambda\,\zeta_{i}(\tau_{i})\,,
\end{equation}
where $m$ is the mass and $\omega$ is the bare frequency of the internal degree of freedom, which we assume to be the same for all detectors. The overall effects of the quantum field show up in  the retarded Green's function $G_{R}^{(\phi)}(x,x')$ of the field and a stochastic forcing term  $\zeta_{i}$  capturing how the field acts on the $i^{\text{th}}$ detector.   For the field state we choose, the classical noise $\zeta$, which represents quantum field fluctuations,  satisfies  Gaussian statistics, determined completely by its second moment in this case,
\begin{align}\label{E:dktuehs}
	\llangle\zeta_{i}(\tau_{i})\rrangle&=0\,,&\llangle\zeta_{i}(\tau_{i})\zeta_{j}(\tau_{j})\rrangle&=G_{H}^{(\phi)}(z^{\mu}_{i},z^{\mu}_{j})\,,
\end{align}
in which $\llangle\cdots\rrangle$ represents taking the stochastic ensemble average whose probability distribution functional can be expressed by the zero mean and the second moment. The two relevant Green  functions of the scalar field, the retarded Green function $G_{R}$ and the Hadamard function $G_{H} $  are defined by
\begin{align}
	G_{R}^{(\phi)}(x,x')&=i\,\theta(t-t')\,\bigl[\phi(x),\phi(x')\bigr]\,,&G_{H}^{(\phi)}(x,x')&=\frac{1}{2}\,\operatorname{Tr}_{\phi}\Bigl(\rho_{\phi}\bigl\{\phi(x),\phi(x')\bigr\}\Bigr)\,,
\end{align}
where the density {operator} $\rho_{\phi}$ is the initial state of the scalar field, and $\theta(t)$ is a unit-step function. We see  that $G_{R}$ is independent of the field state, while $G_{H}$ is state dependent.

The equation of motion \eqref{E:bfsrdf1} is an integro-differential equation, essentially describing a driven oscillator by the scalar field fluctuations. The integral expression contains a local contribution, depending {$Q_{i}$ itself}. It accounts for frequency renormalization  and a dissipative self-force from  the radiative reaction of the field due to the motion of the internal degree of freedom . On the other hand, the nonlocal effect arising from all other detectors  has a causal, history-dependent nature. On account of this variety of backreactions of distinct natures,  how does each of these factors  contribute   to the establishment of  dynamical equilibrium for the totality of the motions of the internal degrees of freedom of all the detectors interacting with the field?  The answer can be found in a generalized FDR, if it does exist.  By examining the dynamical features laden with the equation of motion \eqref{E:bfsrdf1} we shall show that this  {relation} does exist for the system under investigation. We then explore its implications in the FDR and CPR forms.

\subsection{Wightman function stationary in Rindler time}

Before proceeding with the  {dynamical} %dynamics 
analysis, and for future reference, we first show that the Wightman function of the massless scalar field in Minkowski vacuum $\lvert0_{\textsc{m}}\rangle$, when expressed in the Rindler coordinates $(\eta,\mathbf{x}_{\perp},\xi)$, is stationary in Rindler time $\eta$. The field operator in the $R$ wedge is expanded as~\cite{Unr76,HIUY}
\begin{equation}
	\phi(x)=\int_{0}^{\infty}\!d\kappa\int_{-\infty}^{\infty}\!d^{2}k_{\perp}\;\Bigl[\hat{a}_{\kappa\mathbf{k}_{\perp}}^{(\textsc{r})}\,g_{\kappa\mathbf{k}_{\perp}}^{(\textsc{r})}(\eta,\mathbf{x}_{\perp},\xi)+\mathrm{h.c.}\Bigr]
\end{equation}
with $x=(t,\mathbf{x}_{\perp},z)$ in Cartesian coordinates or $x=(\eta,\mathbf{x}_{\perp},\xi)$ in Rindler coordinates. The mode function $g$ is given by
\begin{equation}
	g_{\kappa\mathbf{k}_{\perp}}^{(\textsc{r})}(\eta,\mathbf{x}_{\perp},\xi)=\biggl(\frac{\sinh\frac{\pi\kappa}{\mathsf{a}}}{4\pi^{4}\mathsf{a}}\biggr)^{\frac{1}{2}}\,K_{+i\frac{\kappa}{\mathsf{a}}}(\frac{q}{\mathsf{a}}\,e^{\mathsf{a}\xi})\,e^{+i\mathbf{k}_{\perp}\cdot\mathbf{x}_{\perp}}e^{-i\kappa\eta}\,,
\end{equation}
where $K_{\mu}(z)$ is the modified Bessel function of order $\mu$, and $q=\lvert\mathbf{k}_{\perp}\rvert$ is the magnitude of the transverse momentum $\mathbf{k}_{\perp}$ of the massless field. The creation and annihilation operators $\hat{a}_{\kappa\mathbf{k}_{\perp}}^{(\textsc{r})\dagger}$, $\hat{a}_{\kappa\mathbf{k}_{\perp}}^{(\textsc{r})}$ satisfy the standard commutation relations
\begin{equation}
	\bigl[\hat{a}_{\kappa\mathbf{k}_{\perp}}^{(\textsc{r})},\,\hat{a}_{\kappa'\mathbf{k}'_{\perp}}^{(\textsc{r})\dagger}\bigr]=\delta(\kappa-\kappa')\,\delta^{(2)}(\mathbf{k}_{\perp}-\mathbf{k}'_{\perp})
\end{equation}
and zero otherwise. The superscript $(R)$ signifies that the associated quantities are defined in the right Rindler wedge.

The Wightman function $G_{>}^{(\phi)}(x,x')=i\,\langle0_{\textsc{m}}\vert\phi(x)\phi(x')\vert0_{\textsc{m}}\rangle$ is then given by
\begin{align}
	G_{>}^{(\phi)}(x,x')&=i\int_{0}^{\infty}\!d\kappa\;J_{\kappa}(\xi,\mathbf{x}_{\perp};\xi',\mathbf{x}'_{\perp})\Bigl[\coth\frac{\pi\kappa}{\mathsf{a}}\,\cos\kappa(\eta-\eta')-i\,\sin\kappa(\eta-\eta')\Bigr]\,,\label{E:vfhvgfs}
\end{align}
where we have used the facts that 
\begin{align}
	\langle0_{\textsc{m}}\vert\hat{a}_{\kappa\mathbf{k}_{\perp}\vphantom{\mathbf{k}'_{\perp}}}^{(\textsc{r})\vphantom{\dagger}}\hat{a}_{\kappa'\mathbf{k}'_{\perp}}^{(\textsc{r})\dagger}\vert0_{\textsc{m}}\rangle&=\frac{e^{\frac{2\pi\kappa}{\mathsf{a}}}}{e^{\frac{2\pi\kappa}{\mathsf{a}}}-1}\,\delta(\kappa-\kappa')\delta^{(2)}(\mathbf{k}_{\perp}-\mathbf{k}'_{\perp})\,,\\
	\langle0_{\textsc{m}}\vert\hat{a}_{\kappa\mathbf{k}_{\perp}\vphantom{\mathbf{k}'_{\perp}}}^{(\textsc{r})\dagger}\hat{a}_{\kappa'\mathbf{k}'_{\perp}}^{(\textsc{r})\vphantom{\dagger}}\vert0_{\textsc{m}}\rangle&=\frac{1}{e^{\frac{2\pi\kappa}{\mathsf{a}}}-1}\,\delta(\kappa-\kappa')\delta^{(2)}(\mathbf{k}_{\perp}-\mathbf{k}'_{\perp})\,.
\end{align}
Here we have introduced a spectral-like function $J_{\kappa}(\xi,\mathbf{x}_{\perp};\xi,\mathbf{x}'_{\perp})$ to account for all  the spatial variations of the Wightman function
\begin{equation}
	J_{\kappa}(\xi,\mathbf{x}_{\perp};\xi',\mathbf{x}'_{\perp})=\biggl(\frac{\sinh\frac{\pi\kappa}{\mathsf{a}}}{4\pi^{4}\mathsf{a}}\biggr)\int_{-\infty}^{\infty}\!d^{2}k_{\perp}\;K_{+i\frac{\kappa}{\mathsf{a}}}(\frac{q}{\mathsf{a}}\,e^{\mathsf{a}\xi})K_{+i\frac{\kappa}{\mathsf{a}}}(\frac{q}{\mathsf{a}}\,e^{\mathsf{a}\xi'})\,e^{+i\mathbf{k}_{\perp}\cdot(\mathbf{x}_{\perp}-\mathbf{x}'_{\perp})}\,.
\end{equation}

\subsection{FDR-CPRs in the quantum field and between the detectors}

We see the Wightman function \eqref{E:vfhvgfs} of the scalar field has the desired features that it is translationally invariant in Rindler time $\eta$. Then, in terms of Fourier transforms of the retarded Green's function and the Hadamard function of the field,
\begin{equation}
	\widetilde{G}^{(\phi)}(\kappa)=\int_{-\infty}^{\infty}\!d\eta\;G^{(\phi)}(\eta)\,e^{+i\kappa\eta}\,,
\end{equation}
we come up the standard generalized FDR for the field
\begin{equation}\label{E:itrirndd}
	\widetilde{G}_{H}^{(\phi)}(\kappa)=\coth\frac{\pi\kappa}{\mathsf{a}}\,\operatorname{Im}\widetilde{G}_{R}^{(\phi)}(\kappa)\,,
\end{equation}
the same as has been derived for the (1+1)D case~\cite{CPR2D}.

Translational invariance of the two-point function in Rindler time enables us to write the integro-differential equation into an algebraic one after we perform the Laplace or Fourier transformation on the equation of motion.  For this, we first express the equation of motion \eqref{E:bfsrdf1} in terms of the Rindler time,
\begin{align}\label{E:brvsuyr1}
	&m\,e^{-2\mathsf{a}\xi_{i}}\frac{d^{2}Q_{i}}{d\eta_{i}^{2}}(\eta_{i})+m\omega^{2}\,Q_{i}(\eta_{i})\\
	&\qquad\qquad-\lambda^{2}\sum_{j}\int^{\eta_{i}}_{0}\!d\eta_{j}\;\Bigl\{G_{R}^{(\phi)}[\eta_{i},\mathbf{z}_{i}(\eta_{i});\eta_{j},\mathbf{z}_{j}(\eta_{j})]\Bigr\}_{ij}\,e^{+\mathsf{a}\xi_{j}}\,Q_{j}(\eta_{j})=\lambda\,\zeta_{i}(\eta_{i})\,,\notag
\end{align}
and introduce a matrix form for the retarded Green's function $\bigl[\mathbf{G}^{(\phi)}_{R}(\eta,\eta')\bigr]_{ij}\equiv G^{(\phi)}_{R}[\eta_{i},\mathbf{z}_{i}(\eta_{i});\eta_{j},\mathbf{z}_{j}(\eta_{j})]$. To solve the equation of motion, it proves convenient to introduce $\widetilde{\mathbf{D}}^{(2)}(\kappa)$ by
\begin{align}\label{E:bghfhvd1}
	\widetilde{\mathbf{D}}^{(2)}(\kappa)&=\Bigl[-m\,\kappa^{2}\mathbf{A}^{2}+m\omega^{2}\,\mathbf{I}-\lambda^{2}\widetilde{\mathbf{G}}_{R}^{(\phi)}(\kappa)\cdot\mathbf{A}^{-1}\Bigr]^{-1}\,,
\end{align}
where its $ij$ th component implicitly depends on $\xi_{i}$, $\mathbf{x}_{\perp,i}$ and $\xi_{j}$, $\mathbf{x}_{\perp,j}$. The   matrix elements $A_{ij}=e^{-\mathsf{a}\xi_{i}}\,\delta_{ij}$ describe the redshift factor associated with each trajectory of the detector at fixed $\xi_{i}$,   its physical significance will be clearly seen later. The Fourier transformation of~\eqref{E:bghfhvd1} gives $\mathbf{D}^{(2)}(\eta)$, which belongs to a special set of homogeneous solutions to the equation of motion, satisfying $\mathbf{D}^{(2)}(0)=0$ and $\dot{\mathbf{D}}^{(2)}(0)=1$.

Since the subsequent discussions on energy balance, the FDR and relaxation depend only on time,  with no explicit reference to the spatial coordinates -- their dependence appears only in the matrix indices -- all the derivations and calculations\footnote{Whether the two-point Green's functions is translationally invariant in $\xi$ or not plays no role in the derivations.} done in the $1+1$ case can be carried over here.

The inhomogeneous solution of the internal degree of freedom $Q$ is then given by
\begin{equation}\label{E:brividgerwer}
	Q_{i}^{(\textsc{inh})}(\eta)=\lambda\int^{\eta}_{0}\!d\eta'\;D^{(2)}_{ij}(\eta-\eta')\,\zeta_{j}(\eta')\,.
\end{equation}
The inhomogeneous solutions dictate  the late-time dynamics of $Q$ and thus will impact on the relaxation dynamics  of the system at late times, the pre-condition for the power balance and the existence of the generalized FDR. However, renormalization in the formal expression of $\mathbf{D}^{(2)}$ {has to be dealt with first,} as in the case of inertial detectors in $1+3$ dimensional Minkowski space in~\cite{QTD1,LinHu06} because in the coincident limit of {spatial} coordinates, the commutator of the field operator $G^{(\phi)}(x,x')=i\,\bigl[\phi(x),\phi(x')\bigr]$, independent of the field state, is singular
\begin{align}
	\lim_{\mathbf{x}'\to\mathbf{x}}G^{(\phi)}(x,x')&=-\frac{e^{-2\mathsf{a}\xi}}{2\pi}\frac{\partial}{\partial\eta}\,\delta(\eta-\eta')\,,\label{E:ryubwerw}
\end{align}
due to the fact that
\begin{align}
	J_{\kappa}(\xi,\mathbf{x}_{\perp};\xi,\mathbf{x}_{\perp})=\biggl(\frac{\sinh\frac{\pi\kappa}{\mathsf{a}}}{4\pi^{4}\mathsf{a}}\biggr)\int_{-\infty}^{\infty}\!d^{2}k_{\perp}\;K_{+i\frac{\kappa}{\mathsf{a}}}^{2}(\frac{q}{\mathsf{a}}\,e^{\mathsf{a}\xi})=\frac{e^{-2\mathsf{a}\xi}}{4\pi^{2}}\,\kappa\,.
\end{align}
Thus we decompose the integral expression in \eqref{E:brvsuyr1} into
\begin{align}
	&\quad\sum_{j}\int\!d\eta_{j}\;\Bigl\{G_{R}^{(\phi)}[\eta_{i},\mathbf{z}_{i}(\eta_{i});\eta_{j},\mathbf{z}_{j}(\eta_{j})]\Bigr\}_{ij}e^{+\mathsf{a}\xi_{j}}\,Q_{j}(\eta_{j})\label{E:oitiuererw}\\
	&=\int\!d\eta_{j}\;\Bigl\{G_{R}^{(\phi)}[\eta_{i},\mathbf{z}_{i}(\eta_{i});\eta_{j},\mathbf{z}_{i}(\eta_{j})]\Bigr\}_{ii}e^{+\mathsf{a}\xi_{i}}\,Q_{i}(\eta_{j})\notag\\
	&\qquad\qquad\qquad\qquad\qquad+\sum_{j\neq i}\int\!d\eta_{j}\;\Bigl\{G_{R}^{(\phi)}[\eta_{i},\mathbf{z}_{i}(\eta_{i});\eta_{j},\mathbf{z}_{j}(\eta_{j})]\Bigr\}_{ij}\,e^{+\mathsf{a}\xi_{j}}\,Q_{j}(\eta_{j})\,.\notag
\end{align}
The divergent contribution of the first term on the righthand will be absorbed into natural frequency renormalization and its finite part will contribute to the damping term, while the second term will account for the non-Markovian influence originated from the other detectors. A detailed treatment can be found in~\cite{QTD1,LinHu06,LinHu07}. In short, the presence of damping term in linear dynamics inevitably implies that the late-time dynamics of the linear system is governed by the stochastic noise. The contribution from the initial condition is exponentially suppressed at late times.

If we define the retarded Green's function $\mathbf{G}_{R}^{(Q)}$ of the internal degree of freedom $Q$ by {
\begin{equation}
 \widetilde{\mathbf{G}}_{R}^{(Q)}(\kappa)\equiv \widetilde{\mathbf{D}}^{(2)}(\kappa)\cdot\mathbf{A},
\end{equation}}
such that
\begin{align}
	\widetilde{\mathbf{G}}_{R}^{(Q)}(\kappa)=\Bigl[-m\,\kappa^{2}\mathbf{A}+m\omega^{2}\,\mathbf{A}^{-1}-\lambda^{2}\mathbf{A}^{-1}\cdot\widetilde{\mathbf{G}}_{R}^{(\phi)}(\kappa)\cdot\mathbf{A}^{-1}\Bigr]^{-1}\,,\label{E:ekbhgsds}
\end{align}
or $\bigl[\widetilde{\mathbf{G}}_{R}^{(Q)}(\kappa)\bigr]_{ij}=\bigl[\widetilde{\mathbf{D}}^{(2)}(\kappa)\bigr]_{ij}e^{-\mathsf{a}\xi_{j}}$, then $\mathbf{G}_{H}^{(Q)}(\eta_{i},\eta_{j})$, the anti-commutator of the internal degree of freedom $Q$, will be defined in the standard way\footnote{The righthand side of \eqref{E:gksbdds} does not at first sight look like an expectation value of an anti-commutator. But if we recall that it is actually a  statistical average like the one in \eqref{E:dktuehs},  then from the construction of the stochastic effective action of the reduced system, the righthand side of \eqref{E:dktuehs} is indeed equivalent to the expectation value of an anti-commutator. We can also verify this equivalence by the density operator method, see~\cite{HHAoP} for the details and  discussions.}
\begin{equation}\label{E:gksbdds}
	\Bigl[\mathbf{G}_{H}^{(Q)}(\eta_{i},\eta_{j})\Bigr]_{ij}\equiv \llangle Q_{i}(\eta_{i})Q_{j}(\eta_{j})\rrangle\,,
\end{equation}
Then from \eqref{E:brividgerwer}, we find
\begin{align}\label{E:uryrer}
	&\quad\llangle Q_{i}(\tau_{i})Q_{j}(\tau_{j})\rrangle\\
	&=\lambda^{2}\int_{-\infty}^{\infty}\!\frac{d\kappa}{2\pi}\;\Bigl[\widetilde{\mathbf{G}}_{R}^{(Q)*}(\kappa)\Bigr]_{ik}\,\Bigl[\widetilde{\mathbf{G}}_{R}^{(Q)}(\kappa)\Bigr]_{jl}\,\Bigl[\mathbf{A}^{-1}\cdot\widetilde{G}_{H}^{(\phi)}(\kappa)\cdot\mathbf{A}^{-1}\Bigr]_{kl}\,e^{-i\kappa(\eta_{i}-\eta_{j})}+\cdots\,.\notag
\end{align}
where we have used \eqref{E:brividgerwer} and the fact that $\mathbf{D}^{(2)}(\tau)$ is in fact a retarded kernel of $Q$. We have ignored terms that are exponentially small at late times. 
%The bra-ket pair (used often to denote quantum expectation values) here  represents the ensemble average according to the statistics of the stochastic noise. 
The results in \eqref{E:uryrer} shows that at late time the anti-commutator of $Q$ also becomes translationally-invariant in the Rindler time, and the non-stationary component turns out to be exponentially small. This is one of the essential properties to enable us to show the existence of  the FDR for the internal degree of freedom $Q$ from the  nonequilibrium dynamics of open systems.

We may further re-write \eqref{E:uryrer}. Owing to the identity
\begin{align}
	\operatorname{Im}\Bigl[\widetilde{\mathbf{G}}_{R}^{(Q)}(\kappa)\Bigr]^{-1}&=-\lambda^{2}\,\mathbf{A}^{-1}\cdot\Bigl[\operatorname{Im}\widetilde{\mathbf{G}}_{R}^{(\phi)}(\kappa)\Bigr]\cdot\mathbf{A}^{-1}\,,
\end{align}
Eq.~\eqref{E:uryrer} becomes, with the help of the generalized FDR of the field \eqref{E:itrirndd},
\begin{align}
	\Bigl[\mathbf{G}_{H}^{(Q)}(\eta_{i},\eta_{j})\Bigr]_{ij}&=\int_{-\infty}^{\infty}\!\frac{d\kappa}{2\pi}\;\coth\frac{\pi\kappa}{\mathsf{a}}\,\operatorname{Im}\widetilde{\mathbf{G}}_{R}^{(Q)}(\kappa)\,e^{-i\kappa(\eta_{i}-\eta_{j})}+\cdots\,.
\end{align}
Since the integrand is the Fourier transform of $\mathbf{G}_{H}^{(Q)}(\eta_{i}-\eta_{j})$, we also obtain a generalized FDR for the internal degrees of freedom of the uniformly-accelerating detectors in 1+3 Minkowski space,
\begin{equation}\label{E:ncjsewue}
	\widetilde{\mathbf{G}}_{H}^{(Q)}(\kappa)=\coth\frac{\pi\kappa}{\mathsf{a}}\,\operatorname{Im}\widetilde{\mathbf{G}}_{R}^{(Q)}(\kappa)\,,
\end{equation}
accompanied by the generalized FDR of the field \eqref{E:itrirndd}.
%\begin{equation}\label{E:kgbrss}
%\widetilde{\mathbf{G}}_{H}^{(\phi)}(\kappa)=\coth\frac{\pi\kappa}{\mathsf{a}}\,\operatorname{Im}\widetilde{\mathbf{G}}_{R}^{(\phi)}(\kappa)\,.
%\end{equation}

\subsection{Differences from FDR via LRT} 

Here we see similarity with the conventional FDR based on LRT, the central idea of which hinges on weak external disturbance on the system (detector) about its equilibrium state, which is assumed thermal. The bath essentially has no dynamical role except for maintaining a thermal distribution which restrains the system into a thermal state on average.

In contrast, in the current case, the internal degree of freedom of each detector undergoes nonequilibrium evolution, its final equilibrium state, dictated but not completely determined by the bath dynamics, in general is different from the initial state, and even its intermediate states, before the dynamics reaches relaxation, are far from being in (quasi-)equilibrium. Depending on the coupling strength $\lambda$, the final equilibrium state of the system in general does not take on a canonical form.  However  we see  the emergence of the generalized FDR \eqref{E:ncjsewue} between  the internal degrees of freedom established in the final equilibrium state. The proportionality factor in \eqref{E:ncjsewue} plays a pivotal role in this.

The conventional wisdom of LRT says that the factor depends on the universal temperature associated with the thermal state of the system. Nonetheless, in the current case, comparing with \eqref{E:itrirndd} %kgbrss}, 
we observe that 1) the proportionality factor seems to depend on the initial ``temperature'' of the field at $\eta=0$,  and 2)  the final equilibrium state of the system in general is not a Gibbs state.  Even if one tries to introduce an effective temperature, in general it  depends on all sorts of parameters in the configuration,  which undermines the meaning of a universal temperature. We expect the physics behind the proportionality factor in \eqref{E:ncjsewue} is  different from that based on LRT,  except under unusual circumstances. So why does the FDR obtained from the nonequilibrium dynamics of an open system as described in \cite{CPR2D} come to be the same  as that derived from LRT?

The accidental similarity can be understood as follows. First, in the limit of ultra-weak system-bath coupling, the final equilibrium state of the system, via nonequilibrium evolution, will approach to a Gibbs form~\cite{QTD1}, with a temperature identified as the initial bath temperature. This explains why the equilibrium thermal state of the system takes the Gibbs form, from the viewpoint of nonequilibrium physics. Second, since the late-time dynamics of the system is governed by the environment/bath, the statistical nature is passed on to the system~\cite{CPR2D}. This argument also points out a subtle difference between \eqref{E:ncjsewue} and \eqref{E:itrirndd}, %kgbrss}, 
the latter involves the initial state of the field.

For  {uniformly} accelerated detectors the temperature in the proportionality factor $\coth(\beta\kappa/\mathsf{a})$ is the %Unruh 
{fiducial} temperature,
\begin{equation}
	T\sim\frac{\mathsf{a}}{2\pi}\,.
\end{equation}
However, this temperature is not related to the proper acceleration $\alpha=\mathsf{a}\,e^{-\mathsf{a}\xi}$ of each detector, following a trajectory with constant $\xi$, and thus is different from what would be seen\footnote{In the Appendix of~\cite{CPR2D}, we show that each detector sees its own local temperature, related to its proper acceleration.} in the proper frame of each detector. The latter is {given locally in each detector by the Tolman relation~\cite{Tolman},
\begin{equation}
	T_{\textsc{u}}=\frac{T}{\sqrt{-g^{}_{00}}}=\frac{\mathsf{a}\,e^{-\mathsf{a}\xi}}{2\pi}\,,
\end{equation}
which is the Unruh temperature} %thus 
proportional to the proper acceleration of the detector as a probe. Thus, we emphasize that whereas the detector ``feels'' this local  temperature, it is  the common bookkeeping temperature for all detectors which enter in the generalized FDR \eqref{E:ncjsewue}. Using common arguments based on equilibrium thermal dynamics  can lead to a wrong understanding of the underlying physics.

\subsection{Energy balance condition and the generalized FDR}

The physical meaning of the generalized FDR \eqref{E:ncjsewue} is  most easily seen when we examine the  balance of the energy flow between the detectors and the  {surrounding quantum field. 
The off-diagonal elements of  the generalized FDR %matrix relation 
in\eqref{E:ncjsewue}, called the %correlation-propagation relations (
CPR, first discovered  in~\cite{RHA},  is indispensable in keeping the energy balance in the detector-field system, as we now show.}

Referring to the equation of motion \eqref{E:brvsuyr1}, we see that the power delivered by the stochastic noise $\zeta_{i}$ to the internal degree of freedom $Q_{i}$ of the $i^{\text{th}}$ detector is given by
\begin{align}\label{E:thsdkajnw}
	\mathcal{P}^{(i)}_{\zeta}(\tau_{i})&=\lambda^{2}e^{-\mathsf{a}\xi_{i}}  {\sum_j}
	\int_{0}^{\eta_{i}}\!d\eta_{j}\;\frac{d}{d\eta_{i}}\Bigl[\mathbf{D}^{(2)}(\eta_{i}-\eta_{j})\Bigr]_{ij}\,\llangle\zeta_{i}(\eta_{i})\zeta_{j}(\eta_{j})\rrangle\\
	&=\lambda^{2}e^{-2\mathsf{a}\xi_{i}} {\sum_j}\int_{0}^{\eta_{i}}\!d\eta_{j}\;\frac{d}{d\eta_{i}}\Bigl[\mathbf{G}_{R}^{(Q)}(\eta_{i}-\eta_{j})\Bigr]_{ij}\,\Bigl[\mathbf{A}^{-1}\cdot\mathbf{G}_{H}^{(\phi)}(\eta_{i},\eta_{j})\cdot\mathbf{A}^{-1}\Bigr]_{ij}+\cdots\,,\notag
\end{align}
in the comoving frame of the detector, where a comoving observer records the energy flow into or out of the detector. After the motion of the internal degree of freedom is fully relaxed, in the limit $\eta_{i}\to\infty$, we can simplify \eqref{E:thsdkajnw} to
\begin{align}
	\mathcal{P}^{(i)}_{\zeta}(\infty)&=\lambda^{2}e^{-2\mathsf{a}\xi_{i}} {\sum_j}\int_{-\infty}^{\infty}\!\frac{d\kappa}{2\pi}\;\kappa\,\Bigl[\operatorname{Im}\widetilde{\mathbf{G}}_{R}^{(Q)}(\kappa)\Bigr]_{ij}\,\Bigl[\mathbf{A}^{-1}\cdot\widetilde{\mathbf{G}}_{H}^{(\phi)}(\kappa)\cdot\mathbf{A}^{-1}\Bigr]_{ij}\,,\label{E:uthdfs}
\end{align}
which is already independent of time. Meanwhile, the total power delivered by the dissipation and the causal influence is equivalently\footnote{In principle,  the issue of renormalization enters in  the derivation of the power associated with the dissipation.  See the  discussion in~\cite{QTD1}. A derivation can be found in the Appendix of \cite{CPR2D}.} given by
\begin{align}
	&\quad\mathcal{P}^{(i)}_{\gamma}(\tau_{i})+\mathcal{P}^{(i)}_{c}(\tau_{i})\notag\\
	&=\lambda^{2}\,e^{-2\mathsf{a}\xi_{i}} {\sum_j}\int_{0}^{\eta_{i}}\!d\eta_{j}\;\Bigl[\mathbf{A}^{-1}\cdot\mathbf{G}_{R}^{(\phi)}(\eta_{i}-\eta_{j})\cdot\mathbf{A}^{-1}\Bigr]_{ij}\,\frac{d}{d\eta_{i}}\Bigl[\mathbf{G}_{H}^{(Q)}(\eta_{i},\eta_{j})\Bigr]_{ij}\,.
\end{align}
Thus at late times, we arrive at
\begin{align}
	\mathcal{P}^{(i)}_{\gamma}(\infty)+\mathcal{P}^{(i)}_{c}(\infty)&=-\lambda^{2}\,e^{-2\mathsf{a}\xi_{i}}{\sum_j}\int_{-\infty}^{\infty}\!\frac{d\kappa}{2\pi}\;\kappa\,\Bigl[\mathbf{A}^{-1}\cdot\operatorname{Im}\widetilde{\mathbf{G}}_{R}^{(\phi)}(\kappa)\cdot\mathbf{A}^{-1}\Bigr]_{ij}\,\Bigl[\widetilde{\mathbf{G}}_{H}^{(Q)}(\kappa)\Bigr]_{ij}\,, \label{E:rtbkfgbd}
\end{align}
Comparing \eqref{E:uthdfs} and \eqref{E:rtbkfgbd} and making use of the generalized FDRs \eqref{E:ncjsewue} and \eqref{E:itrirndd} %kgbrss},
we conclude
\begin{equation}
	\mathcal{P}^{(i)}_{\zeta}(\infty)+\mathcal{P}^{(i)}_{\gamma}(\infty)+\mathcal{P}^{(i)}_{c}(\infty)=0\,.
\end{equation}
That is, the generalized FDR guarantees that energy flow of each detector is balanced after relaxation, a condition for the  equilibration in the dynamics of the system  of $N$ uniformly accelerating detectors. This fact also implies that the mechanical energy (the sum of the kinetic energy and the harmonic potential energy) of each oscillator, described by the equation of motion \eqref{E:bfsrdf1}, is a time-independent constant. This offers an explanation from the nonequilibrium dynamics  perspective to the common practice in equilibrium thermodynamics, that one  can simply use the Hamiltonian of the system alone to compute the expectation of energy and other physical quantities  without the need to worry about potential complications due to the interaction between the system and the bath, as would be necessary for strong coupling (see, e.g.,  \cite{LinHu07,QTD1,Entropy}).

Finally, a specially noteworthy point is the role of the non-Markovian effects. From the derivation above, it is clear that in addition to the local damping  force (dissipation) and the stochastic force (noise), the energy balance would not hold without the participation of the non-Markovian, causal influence from the other detectors, and nonlocal correlation between detectors residing in the off-diagonal components of the generalized FDR.  {This} is the function and  significance of the {CPR}.\\

\noindent {\bf Acknowledgments} The authors thank Prof. Chong-Sun Chu,  Director of the National Center for Theoretical Sciences in Hsinchu, Taiwan for his hospitality where many discussions in this paper took place.  
SYL is supported by the Ministry of Science and Technology of Taiwan under Grant No. MOST 106-2112-M-018-002-MY3 and in part by the National Center for Theoretical Sciences, Taiwan. KY is supported by MEXT/JSPS KAKENHI Grant No. 15H05895, No. 16H03977 No. 17K05444, No. 17H06359.


\begin{thebibliography}{999}

\bibitem{Kubo} 
R. Kubo, {\it The fluctuation-dissipation theorem}, Rep. Prog. Phys. {\bf 29} (1966) 255.

\bibitem{FetWal}
A. L. Fetter and J. D. Walecka, {\sl Quantum Theory of Many-particle Systems}  (Courier Corporation, Dover 2003)

\bibitem{Sciama}  
D. W. Sciama, {\it Thermal and quantum Fluctuations in special and general relativity: an Einstein synthesis} in {\sl Centenario di Einstein} (Editrici Giunti Barbera Universitaria), 1979).

\bibitem{Haw75}  
S. W. Hawking, {\it Particle creation by black holes}, Commun. Math. Phys. {\bf 43}, 199 (1975).

\bibitem{CanSci77}
P. Candelas and D. W. Sciama, Irreversible Thermodynamics of Black Holes,  Phys. Rev. Lett. 38, 1372 (1977). Erratum Phys. Rev. Lett. 39, 1640 (1977)

\bibitem{Mot86} 
E. Mottola, {\it Quantum fluctuation-dissipation theorem for general relativity}, Phys. Rev. {\bf D33}, 2136 (1986).

\bibitem{HuSin95}  
B. L. Hu and  S. Sinha, {\it Fluctuation-dissipation relation for semiclassical cosmology}, Phys. Rev. D {\bf 51}, 1587 (1995).

\bibitem{CamVer96}
A. Campos and E. Verdaguer, {\it Stochastic semiclassical equations for weakly inhomogeneous cosmologies}, Phys. Rev. D \textbf{53}, 1927 (1996).

\bibitem{CamHu98}        	
A. Campos and B. L. Hu, {\it Non-equilibrium dynamics of a thermal plasma in a gravitational field}, Phys. Rev. D {\bf 58}, 125021 (1998); [hep-ph/9805485].

\bibitem{Unr76} 
W. G. Unruh, {\it Notes on black-hole evaporation}, Phys. Rev. D {\bf 14}, 870 (1976).

\bibitem{SCD} D. Sciama, P. Candelas and D. Deutsch, Adv. Phys. {\bf 30}, 327 (1981).

\bibitem{Tak} S. Takagi, Prog. Theor. Phys. Suppl. {\bf 88}, 1 (1986).

\bibitem{GinFro} V. L. Ginzburg and V. P. Frolov, Sov. Phys. Usp. {\bf 30}, 1073 (1988).

\bibitem{Grove} P.G. Grove, Class. Quantum Grav. {\bf 3}, 801 (1986).

\bibitem{RSG}
D. J. Raine, D. W. Sciama, and P. Grove, {\it Does a uniformly accelerated quantum oscillator radiate}, Proc. R. Soc. A {\bf 435}, 205 (1991).

\bibitem{Unr92} W. G. Unruh, Phys. Rev. {\bf D 46}, 3271 (1992).

\bibitem{MPB} S. Massar, R. Parentani and R. Brout, Class. Quantum Grav. {\bf 10}, 385 (1993).

\bibitem{RHA} 
A. Raval, B. L. Hu, and J. Anglin, {\it Stochastic theory of accelerated detectors in quantum fields}, Phys. Rev. D \textbf{53}, 7003 (1996).

\bibitem{HuRavCapri}
B.L. Hu, Alpan Raval,  {\it Is there emitted radiation in Unruh effect?}Invited Talk at the Capri Workshop on Quantum Aspects of Beam Physics, Oct. 2000 . Proceedings edited by Pisin Chen. (World-Scientific, Singapore, 2001) [quant-ph/0012134]

\bibitem{JH05}
Philip R. Johnson and B. L. Hu, {\it Unruh Effect in a Uniformly Accelerated Charge:  From quantum fluctuations to classical radiation} 
Foundations of Physics {\bf 35} (2005) 1117-1147 [gr-qc/0501029]

\bibitem{LinHu06}
S.-Y. Lin and B. L. Hu, {\it Accelerated detector-quantum field correlations: From vacuum fluctuations to radiation flux}, Phys. Rev. D {\bf73}, 124018 (2006); [quant-ph/0507054].

\bibitem{HIUY}
A. Higuchi, S. Iso, K. Ueda, and K. Yamamoto, {\it Entanglement of the vacuum between left, right, future, and past: The origin of entanglement-induced quantum radiation}, Phys. Rev. D {\bf96}, 083531 (2017).

\bibitem{RHK97} 
A. Raval, B. L. Hu, and D. Koks, {\it Near-thermal radiation in detectors, mirrors, and black holes: A stochastic approach}, Phys. Rev. D {\bf 55}, 4795 (1997).

\bibitem{HJCapri} B. L. Hu, Philip R. Johnson, Beyond Unruh Effect: Nonequilibrium Quantum Dynamics of Moving Charges,  Invited Talk at the Capri Workshop on Quantum Aspects of Beam Physics, Oct. 2000. Proceedings edited by Pisin Chen. (World-Scientific, Singapore, 2001)   	[quant-ph/0012132].

\bibitem{LinHu07}       
S.-Y. Lin and B. L. Hu,  {\it Backreaction and Unruh Effect: New Insights from exact solutions of uniformly accelerated detectors}, Phys. Rev. D {\bf 76}, 064008 (2007) [gr-qc/0611062].

\bibitem{DLBM}
Jason Doukas, S.-Y. Lin, B. L. Hu and R. Mann, {\it Unruh Effect under Nonequilibrium Conditions: Oscillatory Motion of an Unruh-DeWitt detector},  JHEP {\bf 11 (2013)} 119 [arXiv:1307.4360]   %DOI: 10.1007/JHEP11(2013)119 

\bibitem{Lin17} 
S.-Y. Lin, {\it Quantum radiation by an Unruh-DeWitt detector in oscillatory motion}, JHEP {\bf 11 (2017)} 102 [arXiv:1709.08506].

\bibitem{LCH08}
S.-Y. Lin, C.-H. Chou and B. L. Hu, {\it Disentanglement of two harmonic oscillators in relativistic motion}, Phys. Rev. D {\bf78}, 125025 (2008) [arXiv:0803.3995]. 

\bibitem{HLL}		
B. L. Hu,  S.-Y. Lin, and J. Louko, {\it Entanglement between oscillators in relativistic motion and a quantum field}, Class. Quant. Grav. {\bf 29} 224005 (2012); [arXiv:1205.1328]. 

\bibitem{LinHu08} S.-Y. Lin and B. L. Hu,  {\it Quantum Entanglement,  Recoherence and Information Flow in a 	Particle- Field System: Implications for black hole information issue},  Class. Quant. Grav. (special issue) {\bf 25}, 154004 (2008)  [arXiv:0710.0435].

\bibitem{LinHu09} S.-Y. Lin and B. L. Hu, {\it Temporal and Spatial Dependence of Quantum Entanglement 	from Field Theory Perspective},    Phys. Rev. D {\bf 79}, 085020 (2009) [arXiv:0812.4391].

\bibitem{LinHu00}
S.-Y. Lin and B. L. Hu, {\it Entanglement Creation between Two Causally-Disconnected Objects}, 
Phys. Rev. D {\bf 81}, 045019  (2010) [arXiv:0910.5858]. 

\bibitem{OLMH} David Ostapchuk, S.-Y. Lin, R. Mann and B. L. Hu, {\it Entanglement Dynamics between an inertial and an asymptotically uniformly accelerated detector},  JHEP {\bf 07 (2012)} 072 [arXiv:1108.3377].  %10.1007/JHEP07(2012)072

\bibitem{CPR2D} J.-T. Hsiang, B. L. Hu and S.-Y. Lin, {\it Fluctuation-dissipation and correlation-propagation relations from  the nonequilibrium dynamics of detector-quantum field systems}, submitted to Phys. Rev. D.

\bibitem{DeW79}
B. S. DeWitt, {\it Quantum gravity: the new synthesis} in {\sl General Relativity: an Einstein Centenary Survey}, edited by S. W. Hawking and W. Israel (Cambridge University Press, Cambridge, 1979).

\bibitem{CalHu88}
E. Calzetta and B. L. Hu, {\it Nonequilibrium quantum fields: Closed-time-path effective action, Wigner function, and Boltzmann equation}, Phys. Rev. D {\bf 37}, 2878 (1988).

\bibitem{KadBay62}
L. Kadanoff and G. Baym, {\sl Quantum Statistical Mechanics} (Benjamin, New York, 1962).

\bibitem{KuboBook}
R. Kubo,  M. Toda  and N. Hashitsume, {\sl Statistical Physics II: Nonequilibrium Statistical Mechanics}
(Springer Science \& Business Media, 2012).

\bibitem{CalHu08}
E. Calzetta, and B. L. Hu, {\sl Nonequilibrium Quantum Field Theory} (Cambridge University Press, Cambridge, 2008).

\bibitem{Jackson}	
J. D. Jackson, {\sl Classical Electrodynamics}, 2nd Edition, (Wiley, New York, 1975).

\bibitem{Rohlich}
F. Rohrlich, {\sl Classical Charged Particles}, (Addison Wesley, Reading, MA, 1965).

\bibitem{Yaghjian}
A. D. Yaghjian, {\sl Relativistic Dynamics of a Charged Sphere}, Lectures Notes in Physics, Vol. 11, (Springer-Verlag,
New York/Berlin, 1992).

\bibitem{Roh00}
F. Rohrlich, {\it The self-force and radiation reaction}, Am. J. Phys. {\bf 68}, 1109 (2000).

\bibitem{Spohn}
H. Spohn, {\it The critical manifold of the Lorentz-Dirac equation}, Euro. Phys. Lett. {\bf 50}, 287 (2000).

\bibitem{QTD1}
J.-T. Hsiang and B. L. Hu, {\it Quantum thermodynamics from the nonequilibrium dynamics of open systems - Energy, heat capacity, and the third law}'',  Phys. Rev. E \textbf{97}, 0125135 (2018).

\bibitem{FeyVer63}
R. P. Feynman and F. L. Vernon, {\it The Theory of a general quantum system interacting with a linear dissipative system}, Ann. Phys. \textbf{24}, 118 (1963).

\bibitem{CalLeg83} 
A. O. Caldeira and A. J. Leggett, {\it Path integral approach to quantum Brownian motion}, Physica A {\bf121}, 587 (1983).

\bibitem{GSI88}
H. Grabert, P. Schramm, and G. L. Ingold, {\it Quantum Brownian motion: The functional integral
approach}, Phys. Rept. {\bf168}, 115 (1988).

\bibitem{HPZ92}  
B. L. Hu, J. P. Paz, and Y. Zhang, {\it Quantum Brownian motion in a general environment: Exact master equation with nonlocal dissipation and colored noise}, Phys. Rev. D \textbf{45}, 2843 (1992).

\bibitem{HHAoP}
J.-T. Hsiang and B. L. Hu, {\it Nonequilibrium steady state in open quantum systems: influence action, stochastic equation and power balance}, Ann. of Phys. \textbf{362}, 139 (2015).

\bibitem{Sch61}
J. Schwinger, {\it Brownian motion of a quantum oscillator}, J. Math. Phys. {\bf2} 407 (1961). 

\bibitem{Kel63}
L. V. Keldysh, {\sl Diagram technique for nonequilibrium processes}, Zh. Eksp. Teor. Fiz. {\bf 47} 1515 (1964).

\bibitem{Chou85}
K.-C. Chou, Z.-B. Su, B.-L. Hao and L. Yu, {\it Equilibrium and nonequilibrium formalisms made unified}, Phys. Rept. {\bf118}, 1 (1985).

\bibitem{DeW86} 
B. DeWitt, {\it Effective action for expectation values} in {\sl Quantum Concepts in Space and Time}, ed. R. Penrose and C.J. Isham (Clarendon Press, Oxford, Clarendon, 1986).

\bibitem{Jor86} 
R. D. Jordan, {\it Effective field equations for expectation values}, Phys. Rev. D {\bf 33}, 444 (1986).

\bibitem{Tolman}
R. C. Tolman and P. Ehrenfest, {\it Temperature equilibrium in a static gravitational field}, Phys. Rev. {\bf 36}, 1791 (1930).

\bibitem{Entropy}   
J. T. Hsiang and B. L. Hu, {\it Quantum thermodynamics at strong coupling: operator thermodynamic functions and relations}, Entropy {\bf 20}, 423 (2018); [arXiv:1710.03882].  

\end{thebibliography}
\end{document}